\documentclass[a4paper,12pt]{article}

\usepackage{amsmath}
\usepackage{hyperref}
\usepackage{graphicx}
\usepackage[inline]{enumitem}
\usepackage{cleveref}
\usepackage{panacea}
\usepackage{scalerel}

\usepackage{geometry}
\usepackage[skip=12pt]{caption}

\usepackage[T1]{fontenc}
\usepackage[utf8]{inputenc}
\usepackage[scaled=1.04]{biolinum}

\usepackage{fourier}
\usepackage[cal=stixplain]{mathalpha}
\usepackage[scaled=0.83]{beramono}
\usepackage{microtype}

\usepackage{titlesec}
\titleformat{\section}{\sffamily\Large\bfseries}{\sffamily\Large\bfseries\thesection}{0.5em}{\sffamily\Large\bfseries}
\titleformat{\subsection}{\sffamily\large\bfseries}{\sffamily\large\bfseries\thesubsection}{0.5em}{\sffamily\large\bfseries}

\usepackage[numbers,sort&compress]{natbib}
\usepackage[english]{babel}
\makeatletter
\def\NAT@spacechar{\,}
\makeatother

\hypersetup{
    colorlinks=true,
    allcolors=[rgb]{0.26,0.41,0.88},
}

\usepackage{xspace}

\newcommand{\CLs}{\ensuremath{\text{CL}_s}\xspace}
\newcommand{\given}{\,|\,}
\newcommand{\pvalue}{\text{\textit{p}-value}}
\newcommand{\like}{\mathcal{L}}
\newcommand{\ev}{Z}
\newcommand{\nleft}{\mathopen{}\mathclose\bgroup\left}
\newcommand{\nright}{\aftergroup\egroup\right}
\newcommand{\ngiven}{\,\middle\vert\,}

\usepackage{amssymb}
\usepackage{booktabs}
\usepackage{colortbl}
\usepackage{array}
\newcommand{\PreserveBackslash}[1]{\let\temp=\\#1\let\\=\temp}
\newcolumntype{C}[1]{>{\PreserveBackslash\centering}p{#1}}

\newcommand{\code}{\textsf}

\begin{document}

\thispagestyle{empty}
\renewcommand{\thefootnote}{\fnsymbol{footnote}}

\begin{center} 
{\huge\bf\sffamily The Bayes factor surface for searches for new physics}
\end{center}
\vspace{1mm}
\begin{quote}
\begin{center}
{\begin{large}\textsf{\textbf{Andrew Fowlie}}\end{large}\thanks{\textsf{\linkemail{andrew.fowlie@xjtlu.edu.cn}}}}\\[6mm]
{\textit{Department of Physics, School of Mathematics and Physics, Xi'an Jiaotong-Liverpool University, Suzhou, 215123, China}\\}
\end{center}
\end{quote}
\begin{quote}
The Bayes factor surface is a new way to present results from experimental searches for new physics. Searches are regularly expressed in terms of phenomenological parameters --- such as the mass and cross-section of a weakly interacting massive particle. Bayes factor surfaces indicate the strength of evidence for or against models relative to the background only model in terms of the phenomenological parameters that they predict. They provide a clear and direct measure of evidence, may be easily reinterpreted, but do not depend on choices of prior or parameterization. We demonstrate the Bayes factor surface with examples from dark matter, cosmology, and collider physics.
\end{quote}

\makeatletter
\@thanks
\makeatletter
\renewcommand{\thefootnote}{\arabic{footnote}}
\setcounter{page}{1}
\setcounter{footnote}{0}

\section{Introduction}

Experimental searches for new physics are becoming increasingly sophisticated. Nevertheless, experimental results are typically summarized by just a few numbers and plots that express the results of statistical measurements and tests. These results are used by theorists and phenomenologists to understand which models of new physics and which parameters, e.g.~masses and couplings of new particles, should be considered ruled out by experiment.

Indeed, there are two major goals in statistics --- testing, that is, finding which models are ruled out, and measurement, that is, finding which parameters are ruled out (see e.g., ref.~\cite{Kruschke2017}). Testing and measurement are usually conducted under either the frequentist or Bayesian paradigms of statistics. In frequentist frameworks, one conventionally performs measurement by constructing confidence intervals for an unknown parameter and testing by computing \pvalue{}s. To construct a confidence interval, one must find a procedure that generates intervals that include the true value of the unknown parameter at a guaranteed rate~\cite{neyman1937}. The desired rate is known as the confidence level, and $90\%$ and $95\%$ are common choices. Procedures that guarantee that the rate is exactly the confidence level are known as exact, whereas those that guarantee that the rate at least as great as the confidence level are known as valid. The confidence interval may be supplemented by estimates of the parameter, such as a best-fit value.

To test in frequentist frameworks, one constructs a procedure that wrongly rejects a null hypothesis at a guaranteed rate. These procedures commonly use a \pvalue{},
\begin{equation}
    p = \Pr\nleft(t \ge t^\star \ngiven H_0\nright).
\end{equation}
That is, the probability under the null hypothesis that the test-statistic, $t$, is greater than or equal to that observed, $t^\star$. The \pvalue{} allows us to control the error rate: if we reject the null hypothesis when $p < \alpha$, we wrongly reject the null hypothesis at a rate $\alpha$. There exists, however, a duality between measurement and testing (see e.g., ref.~\cite{kendall1987kendall,Lehmann2022}). Confidence intervals for a parameter can be constructed by testing possible parameter values.

In the Bayesian framework, given data $x$, one can perform measurement by considering the posterior, $p(\theta \given x, M)$, for an unknown parameter $\theta$ in model $M$, and summarize it through moments or percentiles. Indeed, the traditional analogue of a confidence interval would be a credible region~\cite{jaynes1976confidence}. This region, $\mathcal{R}$, contains a specified percentile of posterior probability, e.g., $95\%$,
\begin{equation}
    \int_\mathcal{R} p(\theta \given x, M) \, \text{d}\theta = 0.95
\end{equation}
As for confidence intervals, this requires an ordering rule. Credible regions predate confidence intervals; indeed, Laplace used a similar construction to report certainty in the mass of Saturn in 1810~\cite{laplace1810memoire}.

Testing, on the other hand, may be performed by comparing two models, $M_1$ and $M_0$, through a Bayes factor~\cite{kass1995bayes},
\begin{equation}
    B_{10} \equiv \frac{p(x \given M_1)}{p(x \given M_0)} \equiv  \frac{\ev_1}{\ev_0},
\end{equation}
for observed data $x$. The Bayes factor is a ratio of evidences, $\ev$, and tells us how we must update the relative plausibility of two hypothesis in light of data. As emphasized in ref.~\cite{Cousins:2018tiz}, there is no duality between testing and measurement. We cannot infer Bayes factors from credible regions or vice-versa. Measurement conditions on a particular model; testing compares two models. We summarize the frequentist and Bayesian approaches to testing in \cref{tab:summary}.

\begin{table}[t]
    \centering
    \begin{tabular}{r C{4.5cm} C{4.5cm}}
    \toprule
    Goal & Frequentist & Bayesian\\
    \midrule
    Show ruled out parameters & \cellcolor[gray]{0.8} Plot confidence intervals & Plot credible regions\\
    Show ruled out models & \cellcolor[gray]{0.8} Report \pvalue{} & Report Bayes factor\\
    \bottomrule
    \end{tabular}
    \caption{The frequentist and Bayesian approaches to measurement and testing. The frequentist procedures are dual to each other. Loosely follows ref.~\cite{Kruschke2017}.}
    \label{tab:summary}
\end{table}

The duality between testing and measurement in the frequentist setting means that we should reconsider the Bayesian analogue of a confidence interval. We argue that in common cases confidence intervals are used because they test rather than because they measure. In such cases, we should consider the Bayesian analogue of testing not measurement. This occurs particularly when the results of experiments are expressed in terms of phenomenological parameters. Phenomenological parameters are not directly measurable and nor are they fundamental parameters in any theory --- experimental results are expressed through them and theories are mapped to them. In other words, they are a representation of the experimental results that can be readily compared between experiments and interpreted in any model. 

Whenever confidence intervals (or indeed credible regions) were used to test rather than to measure, and we wish to use a Bayesian method to tackle the same problem, we argue that we should use a Bayesian approach to testing. Specifically, we suggest presenting contours of the Bayes factor --- the Bayes factor surface. For example, theories that make predictions that lie outside a Bayes factor contour at $B = 1 / 100$ must be relatively disfavored by at least a factor $100$. 

It was previously suggested to use contours of the Bayes factor to visualize dependence on hyperparameters~\cite{gram2020} and similar ideas recently appeared independently in mathematical statistics~\cite{Johnson_2023}, psychological science~\cite{Wagenmakers2020,Pawel2023} and searches for gravitational waves~\cite{NANOGrav:2023hvm}. In ref.~\cite{Johnson_2023}, Bayes factors are shown as functions of a root mean square effect size (RMSES). Ref.~\cite{Wagenmakers2020,Pawel2023} introduce support intervals as an alternative to credible regions and confidence intervals --- one-dimensional intervals defined by thresholds of a one-dimensional Bayes factor surface. This was not constructed in the context of searches for new phenomena, and thus compares specific choices of parameter against averaging across a prior. 
Lastly, the NANOGrav Collaboration internally developed a Bayes factor surface --- denoted the $K$-ratio ---  and used it to construct one-dimensional upper limits~\cite{NANOGrav:2023hvm}. The details, properties and justification for this procedure were not, however, discussed so far.

\begin{figure}[t]
    \centering
    \includegraphics[width=0.7\textwidth]{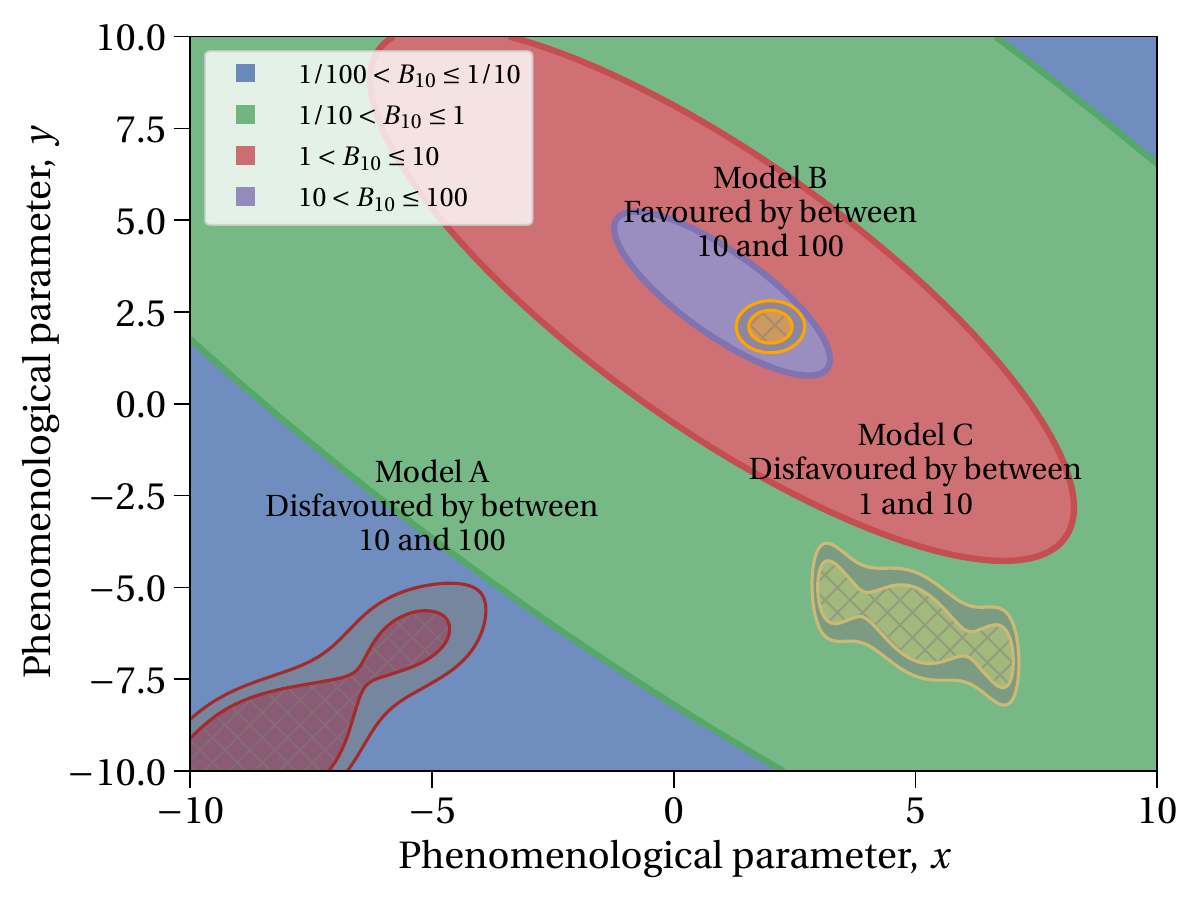}
    \caption{Cartoon of a Bayes factor surface (filled contours). The predictions from three models and the implications are shown.}
    \label{fig:cartoon}
\end{figure}

The rest of this work is structured as follows. In \cref{sec:bf} we explain this proposal and discuss properties and interpretation. In \cref{sec:examples}, we present examples of experimental results in cosmology, astroparticle physics and collider physics that are expressed in phenomenological parameters and used to test.

\section{Bayes factor contours}\label{sec:bf}

\subsection{Definition}

We consider a search for new physics governed by phenomenological parameters $\theta$ and nuisance parameters $\phi$. For reporting and displaying results, we would usually consider at most two phenomenological  parameters. By convention, we assume that no new physics corresponds to $\theta = \theta_0$. We construct a Bayes factor as a ratio of evidences,
\begin{equation}\label{eq:nested}
    B_{10}(\theta) = \frac{\ev(\theta)}{\ev(\theta_0)},
\end{equation}
where the evidence, $\ev$,
\begin{equation}
   \ev(\theta) = \int p(x \given \theta, \phi) p(\phi \given \theta) \, \text{d}\phi,
\end{equation}
for observed data $x$ with likelihood $p(x \given \theta, \phi)$ and prior for the nuisance parameters $p(\phi \given \theta)$. In our Bayes factor, the denominator was the evidence for no new physics, which was nested in our model of new physics. Other choices, including non-nested models, are possible and there may be cases where they are preferred. 

Having constructed the Bayes factor as a function of phenomenological parameters $\theta$, we show results by contours. E.g., the contour of $\theta$ for which $B_{10}(\theta) = 1/10$. Because nuisance parameters are marginalized, the Bayes factor surface can be interpreted as a likelihood ratio for two simple hypotheses. In the nested case, the simple hypotheses are two choices of phenomenological parameters, $\theta$ and $\theta_0$. As the Bayes factor surface corresponds to a likelihood ratio as a function of phenomenological parameters, it does not depend on a choice of prior for the phenomenological parameters. Thus, unlike credible regions, this construction does not depend a choice of prior for the parameters $\theta$, a choice of parameterization or a choice of ordering rule. 

\subsection{Interpretation}

Confidence intervals~\cite{morey2016fallacy} and hypothesis tests~\cite{Greenland2016} may be hard to interpret. On the other hand, the Bayes factor interval admits a simple interpretation: models that make predictions at level $B_{10}$ are at least $B_{10}$ times as plausible relative to the background-only model than they were previously. 

Sometimes we know the posterior predictive for the phenomenological parameters in a new model, $M_{X}$. In these cases, we can estimate the partial Bayes factor, $P_{X0}$, for this new model by averaging $B_{10}(\theta)$, 
\begin{equation}
P_{X0} = \int B_{10}(\theta) p(\theta \given M_X, K) \, \text{d}\theta.
\end{equation}
We assumed here that $p(x \given \theta) = p(x \given \theta, M_X, K)$, that is, that the data, $x$, are conditionally independent of the model, $M$, and any background knowledge, $K$. The partial Bayes factor $P_{X0}$ tells us how to update the relative plausibility of model $M_X$ in light of new data, given the old data. 

Lastly, Bayes factor surfaces for the same models from different experiments are simple to combine: we simply multiply them. For example, to combine Bayes factor surfaces from experiments $E_1$ and $E_2$,
\begin{equation}\label{eq:combine}
    B_{10}(\theta) = B_{10}(\theta; E_1) \times B_{10}(\theta; E_2). 
\end{equation}
This assumes, however, that the experiments $E_1$ and $E_2$ were independent and that there were no common nuisance parameters. 
Suppose  experiments $E_1$ and $E_2$ generate data $x_1$ and $x_2$, respectively. Without loss of generality, the evidence may be written,
\begin{align}
   \ev(\theta) 
   &\equiv \int p(x_1, x_2 \given \theta, \phi) p(\phi \given \theta) \, \text{d}\phi \\
   &= \int p(x_2 \given x_1, \theta, \phi) \, p(x_1 \given \theta, \phi) p(\phi \given \theta) \, \text{d}\phi \\
   &= \int p(x_2 \given x_1, \theta, \phi) \, p(\phi \given \theta, x_1)  \, \text{d}\phi \times Z_1(\theta).
\end{align}
Thus, the posterior for the nuisance parameters from $E_1$ plays the role of the prior for the nuisance parameters in $E_2$.
\Cref{eq:combine} may hold approximately if the experiments are independent and at least one of them depends only weakly on the common nuisances. On the other hand, it could hold exactly if the experiments are independent and if the posterior for the nuisance parameters from $E_1$ was indeed used as the prior in $E_2$.

\subsection{Nuisance parameters and other parameters of interest}

There may be unknown parameters that impact an experiment, but that are not of direct interest. These nuisance parameters should be marginalized, that is, averaged over. These parameters are commonly constrained by background knowledge, e.g., by auxiliary measurements, such that the priors are not controversial.

On the other hand, there may be more than two or three parameters of interest. This number of parameters makes it challenging to visualize a Bayes factor surface or any other type of contour or limit. We could eliminate these parameters by marginalization over a prior. This might, however, make it harder to interpret the resulting contour in a different model and introduces a dependence on the choice of prior. We could instead present slices of parameter space or construct at most two new phenomenological parameters that capture the behavior of the model~\cite{vanBeekveld:2023ney}.

\subsection{Coverage}

Bayes factor contours provide direct measures of evidence; they are not constructed for any particular coverage properties. There are, however guarantees on the properties of the Bayes factor. By Kerridge's theorem~\cite{Kerridge1963},
\begin{equation}\label{eq:kerr}
    \Pr\nleft(B_{10}(\theta) < B \ngiven M_1, \theta\nright) \le B.
\end{equation}
This bound is universal~\cite{Wasserman_2020} --- there are no regularity conditions or asymptotic assumptions. There may, however, be nuisance parameters that were integrated in \cref{eq:kerr},
\begin{equation}
    \Pr(B_{10}(\theta) < B \given M_1, \theta) = \int \Pr(B_{10}(\theta) < B \given M_1, \theta, \phi) \, p(\phi \given M_1, \theta) \, \text{d}\phi.
\end{equation}
Kerridge's theorem thus bounds the expected rate of misleading inferences~\cite{Fowlie:2021zyf}.

A Bayes factor contour at, for example, $B_{10} = 1/10$ would wrongly exclude a parameter at an average rate of no more than $10\%$. If there are no nuisance parameters the average error rate equals the completely frequentist error rate, and thus Bayes factor contours are valid confidence intervals. Because \cref{eq:kerr} only bounds the rate of misleading inferences, Bayes factor contours overcover and are not exact. For example, the long-run coverage of a $B_{10} = 1/10$ interval would be at least $90\%$. Any parameter points that are identical to the null hypothesis would show $B_{10} = 1$ and thus cover at $100\%$. These coverage properties are similar to those of \CLs. 

\subsection{Levels}

In the Neyman-Pearson approach to testing, although we may reject a hypothesis, we cannot quantify evidence for a hypothesis. In the Bayesian framework, we can quantify evidence both for and against a hypothesis. Thus, we may plot contours of Bayes factor at $B_{10} > 1$ and $B_{10} < 1$. The contours at $B_{10} < 1$ show models that are disfavored relative to a chosen alternative; whereas the contours at $B_{10} > 1$ show models that are favored.

We thus suggest reporting several contours of the Bayes factor, perhaps in logarithmic spacing, spanning the range of possible Bayes factors. For example, if on a plane of parameters the Bayes factor spanned $1/100$ to $100$, Bayes factor contours could be shown at $1/100$, $1/10$, $1$, $10$ and $100$. In the absence of evidence for a new theory or effect, the contours at $1/100$ and $1/10$ would show theories that were disfavored or ruled out at those thresholds. In the presence of evidence, contours at $10$ and $100$ would show theories that were favored.



\subsection{Spurious exclusion}

When parameters lie outside a confidence interval, it could indicate either that
\begin{enumerate*}[label=(\roman*)]
    \item the tested parameters are false or
    \item the tested parameters are true and there was a fluctuation away from the predictions of the tested parameters.
\end{enumerate*}
When worrying about spurious exclusion, we worry about the second case, that is, that exclusion indicates a fluctuation. This possibility may be considered disturbing, especially when the test of the parameters was under-powered, that is, $1 - \beta \lesssim \alpha$ for power $\beta$ at error rate $\alpha$~\cite{Highland:1986ee}.

This was a concern in high-energy physics. In this context, the size of a new physics contribution to a measurement can be parameterized by $\mu \ge 0$. The background only model corresponds to $\mu = 0$. Traditional confidence intervals could be found from a choice of test-statistic, $t$, with a \pvalue,
\begin{equation}
    p(\mu) = \Pr\nleft(t > t^\star \ngiven \mu\nright).  
\end{equation}
The parameter value $\mu$ is excluded at $1 - \alpha$ if $p(\mu) < \alpha$. This enables one to find an exact confidence interval. 

There were two proposed modifications to confidence intervals to address this spurious exclusion of $\mu = 0$. First, a \CLs construction~\cite{Read:2000ru,Read:2002hq}. This construction considered the ratio
\begin{equation}
    \CLs(\mu) \equiv \frac{p(\mu)}{p(\mu = 0)}.
\end{equation}
The parameter value $\mu$ is excluded at $1 - \alpha$ if $\CLs(\mu) < \alpha$. Since $\CLs(\mu) \ge p(\mu)$, this results in a valid confidence interval that can never exclude signal models that are sufficiently similar to the background only model. 

Second, a power constrained limit~\cite{Cowan:2011an}. This tackled spurious exclusion by requiring
\begin{equation}
    p(\mu) < \alpha \quad\text{ and }\quad \text{Power}(\mu) > \beta,
\end{equation}
for a chosen threshold $\beta$, where the power,
\begin{equation}
    \text{Power}(\mu^\prime) = \text{Pr}(p(\mu^\prime) < \alpha \given \mu = 0). 
\end{equation}
Whereas $p(\mu)$ depends on the observed data, the power is a property of the experimental design and does not depend on the observed data. Power-constrained limits censure tests of parameter values for which the experiment lacked power.

The spurious exclusion problem is hard to formulate in a Bayesian context. In the Bayesian case, we directly consider the relative plausibility of the tested parameters. That comparison is made through the Bayes factor. In any case, when the tested parameters are identical to the null hypothesis, $B_{10} = 1$ and so the tested parameters cannot lie outside any $B_{10} < 1$ contour.

\subsection{Experimental design}

In a frequentist setting, one would design optimal experiments or analyses by maximizing power~\cite{fisher19711935}. By duality, optimal procedures for measurement are optimal procedures for testing~\cite{kendall1987kendall}. In a Bayesian setting, optimality may be defined through information theory~\cite{Lindley_1956}. That is, optimal experiments maximize the information that we learn. As there is no duality between measurement and testing in the Bayesian framework, however, measurement --- learning about a parameter --- and testing --- learning about models --- are distinct design goals. If we are concerned by testing, e.g.~discovering new phenomena, we may design experiments that are expected to lead to compelling evidence~\cite{Sch_nbrodt_2017}.

\subsection{Computation}

The Bayes factor surface could be computationally challenging, as it may involve challenging evidence integrals~\cite{10.1214/ss/1028905934}. In \cref{eq:nested}, we considered nested models. For nested models, the Bayes factor surface could be found through Savage-Dickey density ratios~(SDDR; \cite{dickey1971weighted}),
\begin{equation}\label{eq:sddr}
    B_{10}(\theta) 
    = \frac{\ev(\theta)}{\ev(\theta_0)}
    = \frac{\ev(\theta) / \ev }{\ev(\theta_0) / \ev}
    = \frac{{p(\theta\given x)}/{\pi(\theta)}}{{p(\theta_0 \given x)}/{\pi(\theta_0)}}
\end{equation}
where
\begin{equation}
    \ev = \int \ev(\theta) \pi(\theta) \, \text{d}\theta.
\end{equation}
As \cref{eq:sddr} can be written in terms of the posterior and prior density, we avoid direct computation of any challenging evidence integrals. Estimating \cref{eq:sddr} through SDDRs requires a choice of prior $\pi(\theta)$, even though ultimately \cref{eq:sddr} should be independent of that choice. We leave discussions of optimal choices of prior for computational efficiency to future works.

There may be a residual challenge; if we want to compute Bayes factor contours at e.g., $B_{10} = 1/100$ through an SDDR, we could require an estimate of the posterior density in the tails of the posterior. In the context of Markov Chain Monte Carlo (MCMC), an accurate estimate of tails could require a substantial effective sample size and thus a long run or multiple chains.
Nested sampling (NS; \cite{Skilling:2006gxv,Ashton:2022grj}) could be an attractive alternative to MCMC, because it compresses through a sequence of constrained priors resulting in weighted samples from the tails of the posterior.

Lastly, non-nested models are even more challenging as we cannot avoid evidence integrals. Using the SDDR, however, we avoid computing the Bayes factor surface across sites on a grid. The Bayes factor surface can be expressed as
\begin{equation}
    B_{10}(\theta) 
    = \frac{\ev(\theta)}{\ev_0} 
    = \frac{\ev(\theta) / \ev}{\ev_0 / \ev}
    = \frac{p(\theta\given x) / \pi(\theta)}{\ev_0 / \ev}
\end{equation}
Thus the numerator can be found through an SDDR. The denominator, however, requires evidence integrals or an estimate of their ratio. The NS algorithm --- or any algorithm that returns posterior and evidence at the same time --- could be used to compute both the SDDR and the evidence $\ev$. A second run could be required for $\ev_0$, if it involves marginalization of nuisance parameters.

\section{Examples}\label{sec:examples}

\subsection{Cosmology --- Planck measurements of the CMB}

The standard model of cosmology posits that our Universe underwent a period of accelerated expansion (see e.g., ref.~\cite{Baumann:2009ds}). This period --- called inflation --- could solve the flatness and horizon problems in cosmology and leave measurable imprints in the cosmic microwave background~(CMB). 

Models of inflation are thus constrained by measurements of the CMB by the Planck experiment~\cite{Planck:2018vyg}. The results are commonly expressed on a plane of phenomenological parameters that impact the CMB: the scalar-to-tensor ratio, $r$, and the spectral tilt, $n_s$. In the Planck presentation, $68\%$ and $95\%$ credible regions are shown on the $(n_s, r)$ plane and predictions from theories of inflation are superimposed. The goal here is testing: testing those fundamental theories against experimental results. 

\begin{figure}[t]
    \centering
    \includegraphics[width=0.7\textwidth]{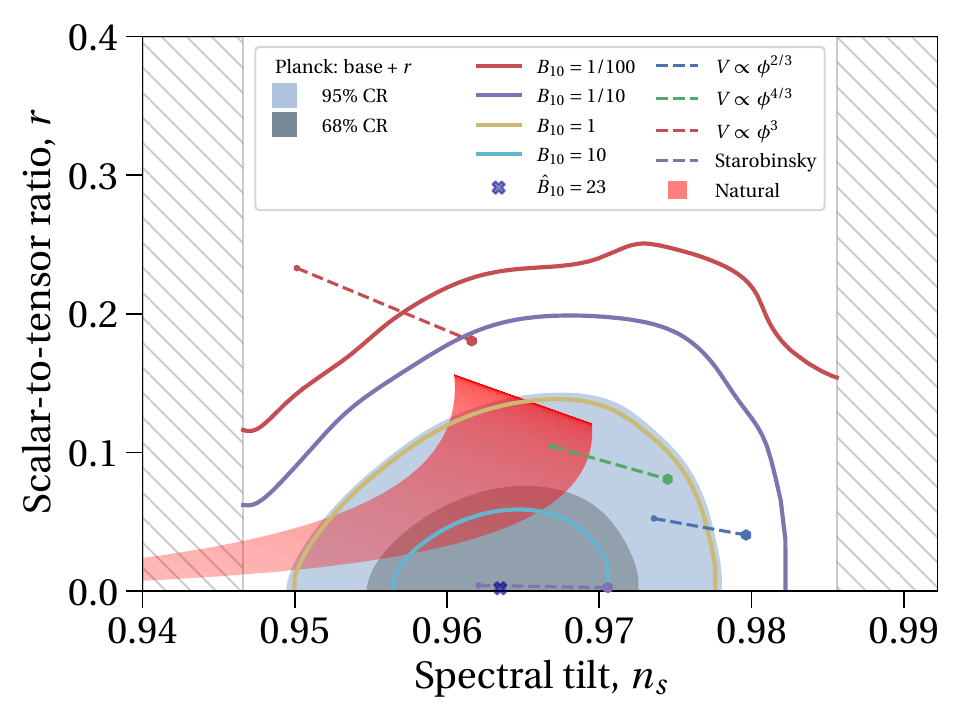}
    \caption{Planck results on the $(n_s, r)$ plane shown through credible regions and a Bayes factor surface relative to the \code{base} model. For reference, we show predictions from several models of inflation.}
    \label{fig:inflation}
\end{figure}

Since the goal is testing, we show instead the Bayes factor surface in \cref{fig:inflation}: we show contours of a Bayes factor for a model that predicts a specific $(n_s, r)$ versus the \code{base} ($r = 0$) model defined by Planck. We computed the surface from Markov Chain Monte Carlo (MCMC) samples produced from \code{cosmoMC}~\cite{Lewis:2002ah} that are available in the Planck legacy archive~\cite{2015scop.confE..36D}.\footnote{For this example we used the \code{base\_r\_plikHM\_TT\_lowl\_lowE\_lensing\_\{1,2,3,4\}.txt} chains  from \code{COM\_CosmoParams\_fullGrid\_R3.01.zip} in ref.~\cite{planck_data}.} Specifically, we computed:
\begin{equation}
    B_{10}(r, n_s) = \frac{\ev_\text{\code{base} + $r$}}{Z_\text{\code{base}}} \frac{\ev(r, n_s)}{\ev_\text{\code{base} + $r$}} = \frac{\pi(r=0)} {p(r=0)}\frac{p(r, n_s)}{\pi(r, n_s)},
\end{equation}
where each factor was expressed through an SDDR. We estimated the posterior densities $p(r=0)$ and $p(r, n_s)$ using kernel density estimation routines from \code{arViz}~\cite{arviz_2019}. There were no MCMC samples in the shaded regions in \cref{fig:inflation} and we do not extrapolate our contours there. Through the SDDR, we marginalize five parameters in the \code{base} + $r$ model (the dark matter and dark energy densities, the Thomson scattering optical depth, the power of the primordial curvature perturbation, and the angular size of sound horizon).

We see in \cref{fig:inflation} that the $95\%$ credible region contour\footnote{For consistency, we recompute credible regions using the same tool-chain as that used for the Bayes factor surface. The credible regions are quantitatively similar to the published ones.} corresponds to a Bayes factor of only around $1$. Models that make predictions inside the $68\%$ credible region, on the other hand, are favored by more than 10. The maximum Bayes factor was $23$. Beyond the $95\%$ credible region, we see contours showing models disfavored by $10$ and $100$. From the superimposed model predictions, we see that Starobinsky inflation may be favored by about $10$ compared to $r = 0$; $2/3$ and $4/3$ mononomial potentials are favored by about $1$ -- $10$ and cubic inflation is disfavored by about 10 or more. Natural inflation makes a range of predictions that sweep across the contours, though only a fraction of predictions are favored by $10$ or more. In summary, in contrast to the credible region, the Bayes factor surface shows us the evidence for or against different models of inflation.

\subsection{Collider physics --- searches for electroweakinos}

Supersymmetry (SUSY) is a popular extension of the Standard Model (SM) of particle physics that predicts a supersymmetric particle for every known SM particle~(see e.g., ref.~\cite{Martin:1997ns}). The fermionic partners of the photon, $W$ boson and Higgs bosons mix to form electroweakinos. There are four neutral electroweakinos, called neutralinos and denoted $\chi^0$, and two charged ones, called charginos and denoted $\chi^\pm$. The predictions of a model depend on the masses and mixing angles of the electroweakinos. Results are typically shown on a two-dimensional plane of phenomenological parameters, using simplifying assumptions for the parameters that are not plotted (though see ref.~\cite{vanBeekveld:2023ney}). These phenomenological parameters are ubiquitous in collider searches and include e.g., signal strength and coupling modifiers in Higgs searches~(see e.g., ref.~\cite{CMS:2022dwd}) and effective field theory coefficients~(see e.g., ref.~\cite{CMS:2023xyc}).

In ref.~\cite{GAMBIT:2018gjo,GAMBIT:2023yih}, a likelihood was constructed for several searches for electroweakinos. To show models that were worse than the background, a capped likelihood ratio versus the background only model was constructed;
\begin{equation}
    \frac{\min\big[\like(s + b), \like(b)\big]}{\like(b)},
\end{equation}
for a model that predicted $s$ signal and $b$ background events, and likelihood $\like$. This was explored by profiling a four-dimensional parameter space to two dimensions. This construction was similar to the Bayes factor surface. 

\begin{figure}[t]
    \centering
    \includegraphics[width=0.7\textwidth]{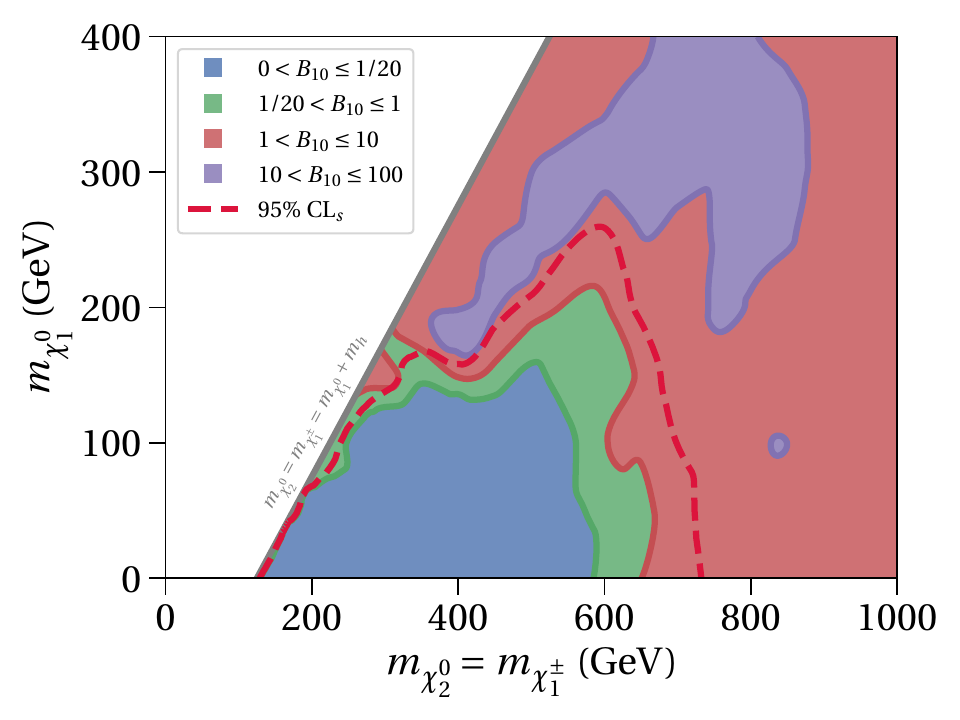}
    \caption{The results from a search for supersymmetric particles with masses $m_{\chi^0_1}$ and $m_{\chi^0_2}$~\cite{ATLAS:2020pgy}. We show a 95\% \CLs exclusion as well as a Bayes factor surface versus the background only model.}
    \label{fig:susy}
\end{figure}

The ATLAS experiment at the Large Hadron Collider (LHC) searched for neutralinos and charginos that were produced from proton collisions and decayed through,
\begin{equation}
    pp \to \chi^{\pm\vphantom{0}}_1 \chi^{0\vphantom{\pm}}_2 \to W^\pm h \chi^0_1 \chi^0_1.
\end{equation}
The Higgs and $W$ were assumed to decay to a pair of $b$-quarks and leptonically, respectively~\cite{ATLAS:2020pgy}. The observed events were consistent with the background only model and $95\%$ \CLs limits were placed on masses of the electroweakinos, assuming that the lightest chargino and next-to-lightest neutralino are mass-degenerate winos, $m_{\chi^\pm_1} = m_{\chi^0_2}$, and that the lightest neutralino is a bino.

We reanalyze this search using \code{pyhf}~\cite{pyhf,pyhf_joss} and the publicly available \code{histfactory} models for this search~\cite{hepdata.90607.v3/r3}. In \cref{fig:susy} we show the Bayes factor surface as well as the $95\%$ \CLs limit on the $(m_{\chi^\pm_1} = m_{\chi^0_2}, m_{\chi^0_1})$ plane. The Bayes factor surface shows the Bayes factor for a specific choice of masses versus the background only model. We see that a chunk of the plane is excluded at $95\%$ \CLs. The Bayes factor surface reveals that this includes models, however, that are favored versus the background only model, $B_{10} > 1$. This discrepancy occurs because \CLs is ratio of tail probabilities whereas the Bayes factor is a ratio of probability densities of the observed data. The contours indicate regions of masses are disfavored by Bayes factors of less than one and less than $20$. Lastly, excesses in the observed data of around $2\sigma$ (local) lead to a region that is favored by between $10$ and $100$; the existence of this region or the strength of the evidence was not revealed by the \CLs contour.

There are $125$ nuisance parameters in this \code{histfactory} model, and it is suggested that $124$ of them are allowed to vary. Marginalizing so many nuisance parameters would be computationally expensive and there is no specialized functionality for this in \code{pyhf}.\footnote{Though see ongoing developments~\cite{Feickert:2023hhr} that enable Hamiltonian Monte Carlo~\cite{Neal1996} on \code{histfactory} models.} We thus fix all nuisance parameters to their central values and compute the Bayes factor surface as a likelihood ratio. For consistency, we fix all nuisance parameters in the \CLs contour. These contours should thus be considered somewhat approximate, though the \CLs contour is quantitatively similar to that in ref.~\cite{ATLAS:2020pgy}.

\subsection{Astroparticle physics --- direct detection of dark matter}

\begin{figure}[t]
    \centering
    \includegraphics[width=0.7\textwidth]{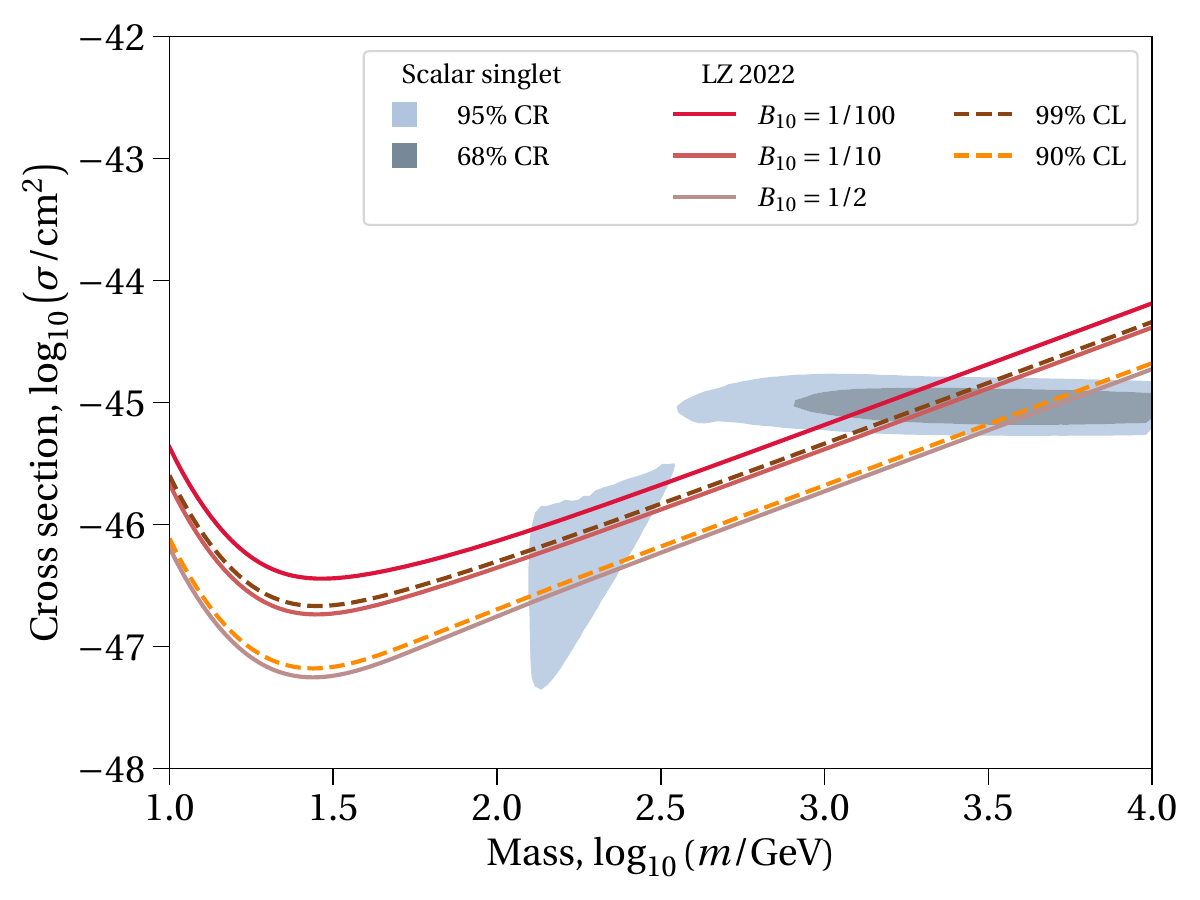}
    \caption{Results from LZ 2022~\cite{LZ:2022lsv} shown through confidence intervals and the Bayes factor surface versus the background only model. Predictions from the scalar singlet model are shown for comparison.}
    \label{fig:direct_detection}
\end{figure}

Weakly Interacting Massive Particles (WIMPs) are popular candidates for dark matter (DM) that could be detected through elastic scattering in direct detection (DD) experiments~\cite{Goodman:1984dc,Jungman:1995df}. The signal depends, among other things, on the unknown mass, $m$ and scattering cross section, $\sigma$, of the WIMP. Thus, results of DD experiments are presented on the phenomenological $(m, \sigma)$ plane. The results are summarized by contours on this plane, traditionally through a confidence interval. These intervals are not uniquely defined by the statistical framework and depend on choices and conventions that differ between experiments. Because of perceived deficiencies in the simplest approaches to setting confidence limits, limits are often set by following specialized techniques including power constrained limits~\cite{Baxter:2021pqo}. 

\begin{figure}[t]
    \centering
    \includegraphics[width=0.9\textwidth]{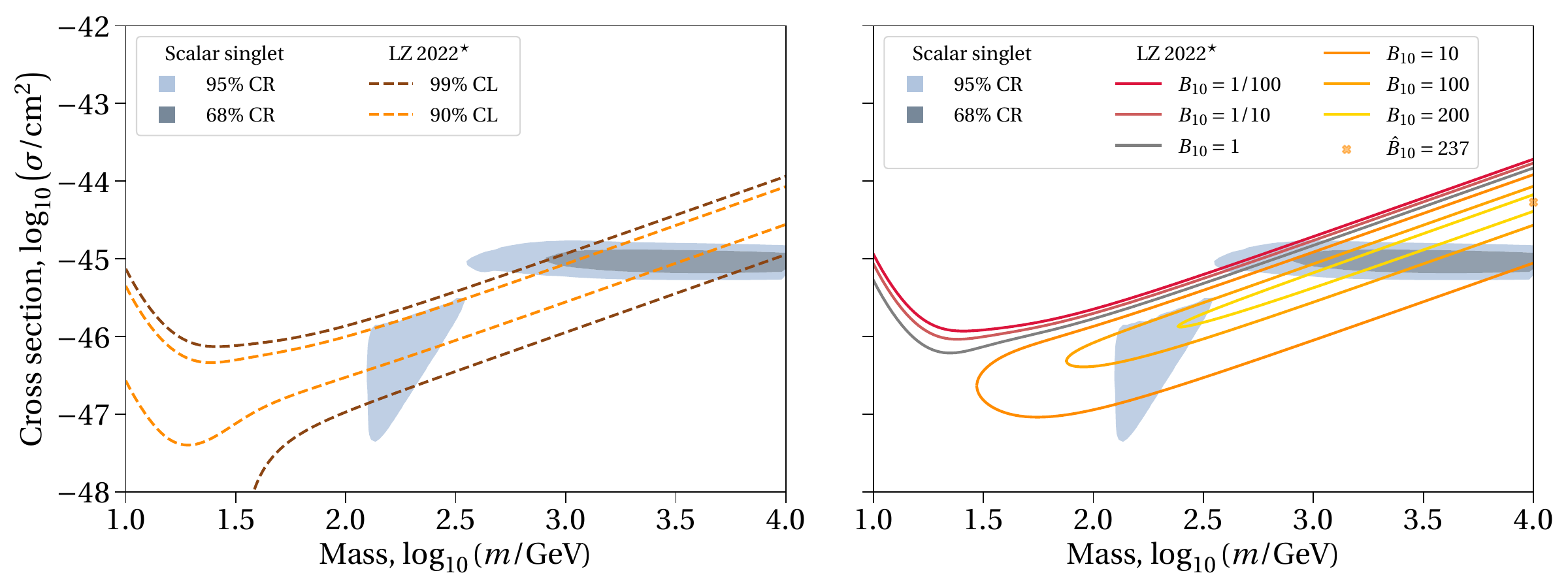}
    \caption{Signal injected into LZ 2022~\cite{LZ:2022lsv} shown through confidence intervals (left) and a Bayes factor surface (right). Predictions from the scalar singlet model are shown for comparison.}
    \label{fig:direct_detection_combined}
\end{figure}

In \cref{fig:direct_detection} we show the Bayes factor surface versus the background only model and one-dimensional confidence limits from the LZ 2022 DD experiment~\cite{LZ:2022lsv} on the $(m, \sigma)$ plane. They were computed using the likelihood implemented in \code{DDCalc}~\cite{GAMBITDarkMatterWorkgroup:2017fax}. We follow the convention that astrophysical nuisance parameters are fixed to benchmark values rather than varied~\cite{Baxter:2021pqo} such that the Bayes factor surface reduces to a likelihood ratio. We compute one-dimensional confidence limits using an asymptotic approximation and they are numerically similar to the official confidence limits.  For comparison, we show predictions from a scalar singlet model of WIMP dark matter~\cite{GAMBIT:2017gge} found from publicly available MCMC chains~\cite{the_gambit_collaboration_2017_846860}. These are posterior predictions; they take into account constraints from previous dark matter searches.

The Bayes factor contours on the $(m, \sigma)$ plane versus the background only model are similar shapes to the confidence limits. They show, however, that the $99\%$ and $90\%$ confidence limits correspond to Bayes factors of only about $2$ and $10$. The contours intersect the predictions from scalar singlet dark matter and indicate the extent to which the model was disfavored by the experimental result. The strength of evidence cannot be deduced from the confidence limit.

In \cref{fig:direct_detection_combined} we show how the contours would look if the LZ experiment had observed a signal of dark matter. We injected $3 \sqrt{b}$ events into the observed events in each bin. By construction, the one-dimensional confidence limits permit the entire mass range. Ref.~\cite{Baxter:2021pqo} suggests that in the presence of a signal two-dimensional limits could be drawn as well. The Bayes factor contours show the extent to which models that predicted cross sections that were too large would be disfavored, as well as the evidence for models that reproduce the injected signal. The maximum Bayes factor is about $237$, lying at a WIMP mass of $10\,\text{TeV}$. Confidence limits, on the other hand, cannot indicate the strength of evidence in favor of a signal.

\section{Conclusions}

We presented the Bayes factor surface --- a new way to present experimental searches for new physics. The results of these searches are commonly expressed in terms of phenomenological parameters. We argued that measuring phenomenological parameters themselves is not of direct interest; rather, we are interested in testing models of new physics. In frequentist paradigms, this point is moot as there is a duality between testing and measurement.

In Bayesian statistics, there is no such duality. We thus argued that the Bayesian analogue of frequentist confidence intervals need not be a credible region or a summary based on the posterior. Rather, if we are interested in the testing aspect of confidence intervals, the Bayesian analogue should be built from a Bayes factor. This led us to the Bayes factor surface. The Bayes factor surface shows the strength of evidence for or against a model relative to the background only model in terms of the phenomenological parameters that it predicts. We demonstrated the Bayes factor surface with examples from Planck measurements of the cosmic microwave background, a collider search for supersymmetric particles, and a direct search dark matter.

The Bayes factor surface provides a clear and direct measure of evidence, may be easily reinterpreted, but does not depend on choices of prior or parameterization. 

\subsection*{Acknowledgments}

We thank Valen Johnson, Kai Schmitz, Samuel Pawel and Ken Olum for discussions.

\bibliographystyle{JHEP}
\bibliography{sample}

\end{document}